\documentclass[showpacs,pra,twocolumn]{revtex4}

\usepackage{epsfig,amsmath}
\usepackage{subfigure}
\usepackage{graphicx}
\usepackage{dcolumn}
\usepackage{stmaryrd}
\usepackage{mathrsfs}
\usepackage{pifont}
\usepackage{amsthm}
\usepackage{amssymb}
\usepackage{bm}
\usepackage{latexsym}
\usepackage{hyperref}
\usepackage{color}
\usepackage{appendix}
\usepackage{braket}
\usepackage{amsmath}

\newcommand{\beq}{\begin{equation}}
	\newcommand{\eeq}{\end{equation}}
\newcommand{\beqa}{\begin{eqnarray}}
	\newcommand{\eeqa}{\end{eqnarray}}

\begin{document}
	
	\title{Time series prediction of open quantum system dynamics}
	
	\begin{abstract}
	Time series prediction (TSP) has been widely used in various fields, such as life sciences and finance, to forecast future trends based on historical data. However, to date, there has been relatively little research conducted on the TSP for quantum physics. In this paper, we explore the potential application of TSP in forecasting the dynamical evolution of open quantum systems. We employ deep learning techniques to train a TSP model and evaluate its performance by comparison with exact solution. We use the ratio of the prediction step length and the sequence length to define short and long-term forecasting. Our results show that the trained model has the ability to effectively capture the inherent characteristics of time series for both short-term and long-term forecasting.  Accurate predictions for different coupling intensities and initial states are obtained. Furthermore, we use our method to train another model and find that it can successfully predict the steady state of the system. These findings suggests that TSP is a valuable tool for the prediction of the dynamics in open quantum systems.

	\end{abstract}
	
	\author{Zhao-Wei Wang$^{1}$, and Zhao-Ming Wang$^{1}$\footnote{wangzhaoming@ouc.edu.cn}}
	\affiliation{$^{1}$ College of Physics and Optoelectronic Engineering, Ocean University of China, Qingdao 266100, China}
	\maketitle
	
	\section{Introduction.}

	The study of an open quantum system dynamics is of fundamental interest for their exploitation in lots of fields including quantum information and quantum optics \cite{verstraete2009quantum,barreiro2011open}. The Lindblad equation has been employed to solve the dynamics of the system which has interactions with its environment \cite{blais2021circuit,marino2016driven,wang2018adiabatic}. Normally, it is a daunting task to solve the dynamics of the system in an environment, especially for a non-Markovian environment \cite{wang2023hybrid,rota2018dynamical}. Many approaches have been developed to tackle this difficulty, such as renormalization group \cite{rota2017critical,finazzi2015corner}, mean field \cite{biella2018linked,jin2016cluster} or quantum state diffusion equation \cite{wang2018adiabatic}. Additionally, machine learning has been applied for studying the dynamics of open quantum systems in recent years \cite{olivera2023benefits,mellak2023quantum,martyn2023variational,chen2022learning,zanoci2023thermalization,wang2023simulation,zhang2023transformer}. It has been proven that deep neural network can efficiently represent most physical states, including the ground states of many-body Hamiltonians and states generated by quantum dynamics \cite{gao2017efficient}. Luo \emph{et al.} introduce a general method that utilizes autoregressive neural networks for simulating open quantum systems \cite{luo2022autoregressive}. The dynamics of open quantum many-body systems can also be simulated using deep autoregressive neural networks through variational methods \cite{reh2021time}. Variational methods based on variational Monte Carlo method and neural network representation based on density matrix can also be used to effectively simulate the non-equilibrium steady state of the Markovian open quantum systems \cite{nagy2019variational}. Another notable nonlinear machine learning model, the Kernel Ridge Regressor, has also been successfully applied to predict the time evolution of quantum systems \cite{rodriguez2022comparative}.

	TSP is a widely utilized method for analyzing past data characteristics to infer future behavior \cite{han2019review, zhou2021informer, chatfield2000time, lim2021time, torres2021deep, masini2023machine}. It finds applications in various domains, including weather forecasting \cite{soares2018ensemble}, stock market trend forecasting \cite{yadav2020optimizing, yan2018financial}, solar power generation forecasting \cite{khan2023quantum}, and biological sequence data analysis \cite{winther2015convolutional}. Moreover, quantum physics plays a significant role in TSP \cite{singh2021fqtsfm, kaushik2022one, emmanoulopoulos2022quantum, mujal2023time, kutvonen2020optimizing, cuellar2023time}. These strategies, repeating part of the experiment after each projection measurement or employing weak measurements, can enable efficient time series processing, and pave the way for different quantum techniques \cite{mujal2023time}. Kutvonen \emph{et al.} investigate the storage capacity and accuracy of quantum library computers based on the fully connected transverse field Ising model, with a particular emphasis on their potential for efficient and accurate TSPs, including the prediction of stock values \cite{kutvonen2020optimizing}. Additionally, using variable component subcircuits as quantum analogies to feedforward artificial neural networks for TSP tasks can improve the results of false predictions while maintaining a smaller number of parameters than the classical machine learning \cite{cuellar2023time}.

	From the above discussion, we notice that TSP is most suitable for the behavior prediction of time evolution, thus it may be applicable for the prediction of open quantum system dynamics. In this paper, we propose a TSP strategy for open quantum system dynamics that utilizes deep learning techniques. By leveraging the Positive-Operator Valued Measure (POVM) method to transform the probability distribution data of the reduced density matrix, we can effectively use well-established machine learning techniques \cite{carrasquilla2019reconstructing} to train the model. These trained models have the ability to capture the past evolution patterns of open quantum systems and provide accurate short-term predictions under various initial states and coupling intensities. Additionally, we try the long-term predictions, i.e., using full prediction datas to make further predictions. Surprisingly, we find that the predicted dynamics is almost the same with the exact solution. We then use the long-term TSP to train another model to predict the steady state of the system and again successful predictions have been obtained.

	\section{FORMALISM}

	For a simple way of describing the open system, Markovian approximation is commonly used, where the memory effects due to system-environment correlations are neglected. In this case, Markovian master equation in Lindblad form is
	\begin{equation}
		\dot{\rho}=-i[H,\rho]+\sum_{k}\frac{\gamma_{k}}{2}\left(2 L_{k} \rho L_{k}^{\dagger}-\{\rho,L_{k}^{\dagger}L_{k}\}\right),
		\label{equ:1}
	\end{equation}
	where $H$ is the system Hamiltonian and $\rho$ is the reduced density matrix of the system, which follows the requirements of the probability conservation and complete positivity of the dynamical map \cite{gorini1976completely}. $\gamma_k$ is the coupling strength between the \emph{k}th mode of the environment and the system. $L_k$ is the Lindblad operator. In order to use the mature machine learning technique, we first apply the POVM to transform the reduced density matrix $\rho$ into one-dimensional probability distribution data \cite{carrasquilla2021probabilistic}. Then the Lindblad equation can be written in the form under probability distributions $P(\textbf{a})$, where $\textbf{a}$ represents a string of measurement operators acting on different qubits. Given an information complete POVM, the probability distribution $P(\textbf{a})$ can be uniquely mapped to the reduced density matrix $\rho$ of the N-spin system
	\begin{equation}
		P(\textbf{a})=\mathrm{tr}(\rho M(\textbf{a})),
		\label{equ:2}
	\end{equation}
	where $M(\mathbf{a}) = M(a_1)\otimes..\otimes M(a_N)$ is one of the positive semidefinite operators $\{M{(\boldsymbol{a})}\}$.  In this paper, the Tetrahedral POVM $ \boldsymbol{M_{\mathrm{tetra}}} = \{M(a) = (1/4)(\mathbb{I} + \textbf{n}^{(a)}\cdot\boldsymbol{\sigma})\} $ has been used. Measurement bases of tetrahedral POVM form a tetrahedral shape on the Bloch sphere, the four vectors $\textbf{n}^{(a)}$ are $ \textbf{n}^{(0)} = (0, 0, 1) $, $\textbf{n}^{(1)} = (2{\sqrt2}/3, 0, -1/3) $, $ \textbf{n}^{(2)} = (-{\sqrt2}/3, -{\sqrt6}/3, -1/3) $ and
	$\textbf{n}^{(3)} = (-{\sqrt2}/3, -{\sqrt6}/3, -1/3) $.

	By inverting Eq.~(\ref{equ:2}), we obtain 
	\begin{equation}
		\rho=\sum_{\mathbf{a}}\sum_{\mathbf{a}^{\prime}}P(\mathbf{a})T^{-1}_{\mathbf{aa}^{\prime}}M({\mathbf{a}^{\prime}}),
		\label{equ:3}
	\end{equation}
	where $T_{\mathbf{aa}^{\prime}} = \operatorname{tr}(M(\mathbf{a})M({\mathbf{a}^{\prime}}))$ is the element of the overlap matrix $T$. In the probability distribution frame, the expectation value of the operator $O$ 
	\begin{equation}
		\langle O\rangle=\sum_{\mathbf{a}}\sum_{\mathbf{a}^{\prime}}P(\mathbf{a})\mathrm{Tr}\left(OM{(\mathbf{a}^{\prime})} T_{\mathbf{a}^{\prime}\mathbf{a}}^{-1}\right).
		\label{equ:4}
	\end{equation}

	\section{TSP AND DEEP LEARNING MODEL}

\begin{figure}
		\includegraphics[width=\linewidth]{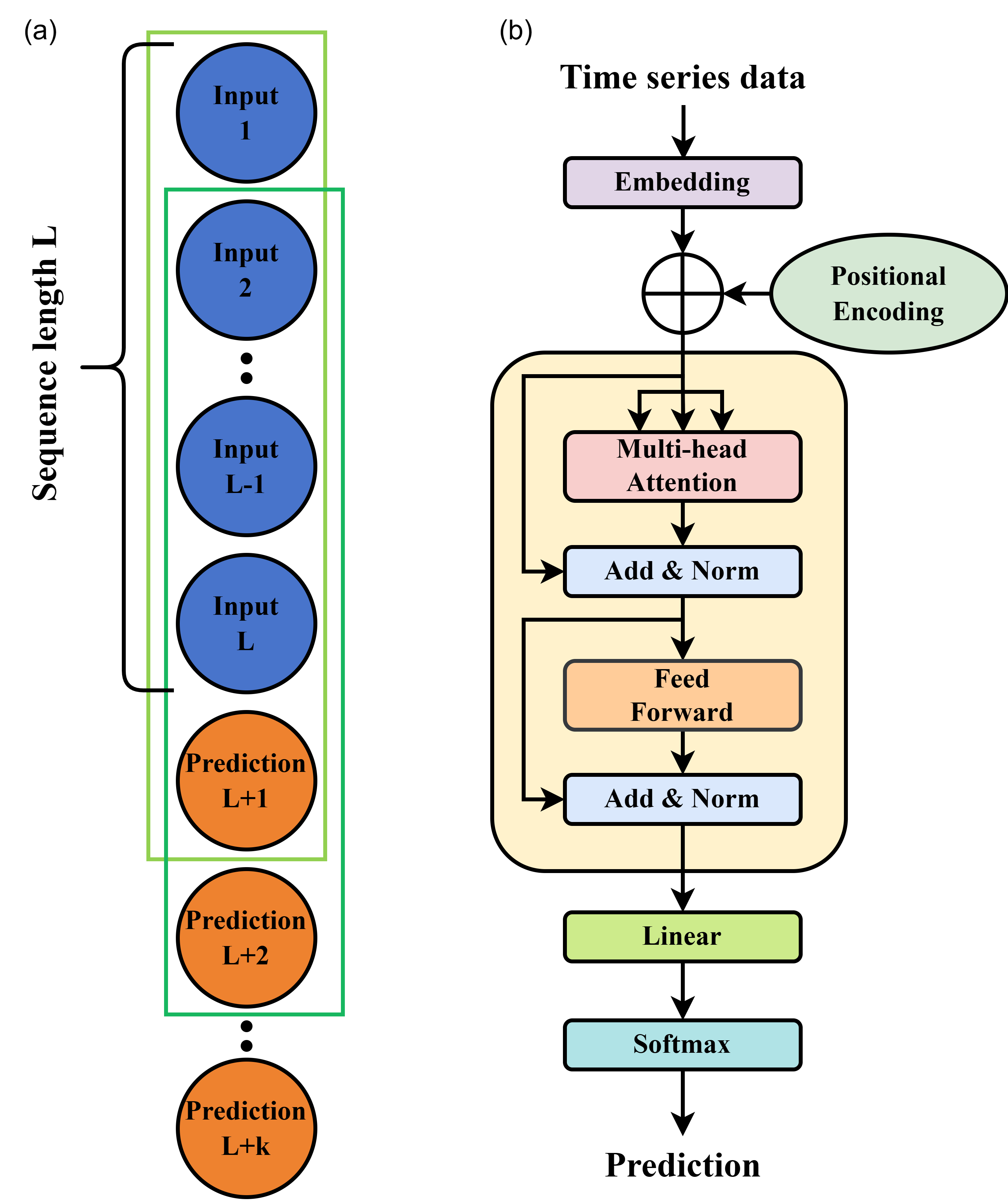}
		\caption{(a) Schematic of the TSP. The blue (orange) ball represents the known (predicted) data. A rectangle represents a TSP unit, and $L$ is the length of the series. The task is to predict the future $(L+1)th$ data from $L$ known data in the past. (b) Schematic of the Transformer model. The time series data is extended by adding positional Encoding to the Embedding layer as input to the Transformer model. After the data is processed by the Transformer model. The output is obtained through the Linear layer and Softmax layer.}
		\label{fig:1}
	\end{figure}

	TSP is a method of forecasting future values based on the characteristics and trends observed in past time series data. It utilizes statistics, machine learning, and deep learning techniques to build predictive models. Fig.~\ref{fig:1}(a) presents a schematic diagram of TSP. The rectangular box represents a prediction action unit, where the time series data with a sequence length of $L$ serves as the basis for the prediction, and the $L + 1$ data point represents the result predicted by the model. To construct a new prediction basis, the $L + 1$ data point is added to the end of the time series, and the first data point is eliminated. This process can be repeated to forecast data for a specific time period in the future. However, it is important to note that this process cannot continue indefinitely. As the forecasting process progresses, an increasing number of predicted data points are utilized to build the forecasting basis. Since the predicted data inherently contains errors compared to the real data, the errors in the later predicted data points will also accumulate.The prediction step whose length is shorter (longer) than the sequence length $L$ is defined as short-term (long-term) TSP.

	Our deep learning model utilizes the Transformer neural network in PyTorch  \cite{paszke2019pytorch} to build an efficient architecture for modeling sequence data. The Transformer model introduces self-attention mechanism and location coding \cite{vaswani2017attention} to enable effective representation and modeling of the input time series data. Fig.~\ref{fig:1}(b) illustrates a schematic diagram of the Transformer model. The input is an autocorrelation matrix consisting of rows of different measurements and columns of measurements at different times. The Embedding layer can make a linear transformation of the time series matrix dimension $d$ and extend to the hidden dimension $d_{model}$. Positional Encoding can be computed by the fixed Eqs.~(\ref{equ:5}-\ref{equ:6}) \cite{vaswani2017attention},

	\begin{equation}
		PE_{p,i}=sin(p/10000^{2i/d_{model}}),
		\label{equ:5}
	\end{equation}
	
	\begin{equation}
		PE_{p,i}=cos(p/10000^{2i/d_{model}}).
		\label{equ:6}
	\end{equation}

	Since the value does not change, and the Positional Encoding matrix can be simply added to the time series matrix at time $t$ if there is any data available. $p$ represents the number of the time series data bars, $i$ represents the number of hidden dimensions $d_{model}$. For odd (even) $i$, the Position Encoding can be  calculated by Eq.~(\ref{equ:5}) (Eq.~(\ref{equ:6})). The computation at the Encoder layer is parallel, and adding positional coding preserves the time relationship of the time series.

	The main principle of the Self Attention Mechanism is to calculate the attention score of the first moment and every moment in the time series (including the first moment), then multiply the calculated attention score by the information of the corresponding moment, and then add together. The result is the weighted sum of the first moment and all the moments in the time series. Finally, the attention information of each moment and time series is updated in turn. More precisely, we multiply the matrix $A$ after Embedding and Positional Encoding by $W^Q$, $W^K$ and $W^V$ respectively to get a query matrix $Q$, a key matrix $K$, and value matrix $V$. Then, the attention value is calculated by Eq.~(\ref{equ:7}) \cite{vaswani2017attention},
	\begin{equation}
		B(Q,K,V)=\operatorname{softmax}\bigg(\frac{QK^\mathrm{T}}{\sqrt{d}}\bigg)V,
		\label{equ:7}
	\end{equation}
	where $\operatorname{softmax}(x_i)=\frac{e^{x_i}}{\sum_je^{x_j}}$. The difference between Multi-Head Attention and Single-Head Attention is that the original three large matrices $Q$, $K$ and $V$ are divided into eight small matrices with the same shape (split in the feature dimension), that is, Eight-Head Attention. The results of each small matrix calculation are then spliced together to get the same matrix $B$ as the results of the Single-Head Attention calculation.

	The Add layer is based on the concept of a residual neural network, where the input matrix $A$ of the Multi-Head Attention is added directly to the output matrix $B$ of the Multi-Head Attention, resulting in the sum matrix $B^{\prime}$. Subsequently, Layer Normalization is applied, which normalizes each row of $B^{\prime}$ to follow a standard normal distribution, yielding the final result $B^{{\prime}{\prime}}$. The Feed Forward layer consists of two fully connected layers, with a ReLU activation function \cite{Goodfellow-et-al-2016} sandwiched between them. At last, the output results are obtained by passing the data through a Linear layer followed by a Softmax activation layer.

	During training, we employed an Encoder layer and Decoder layer and extended the data to $d_{model}= 32$. The Mean Squared Error loss function is used, and the model parameters are updated using the Adam optimizer \cite{kingma2014adam} with a learning rate 0.00001.

	\section{RESULTS AND DISCUSSIONS}
	\subsection{Short-term TSP}

	\begin{figure}
		\centerline{\includegraphics[width=1.0\columnwidth]{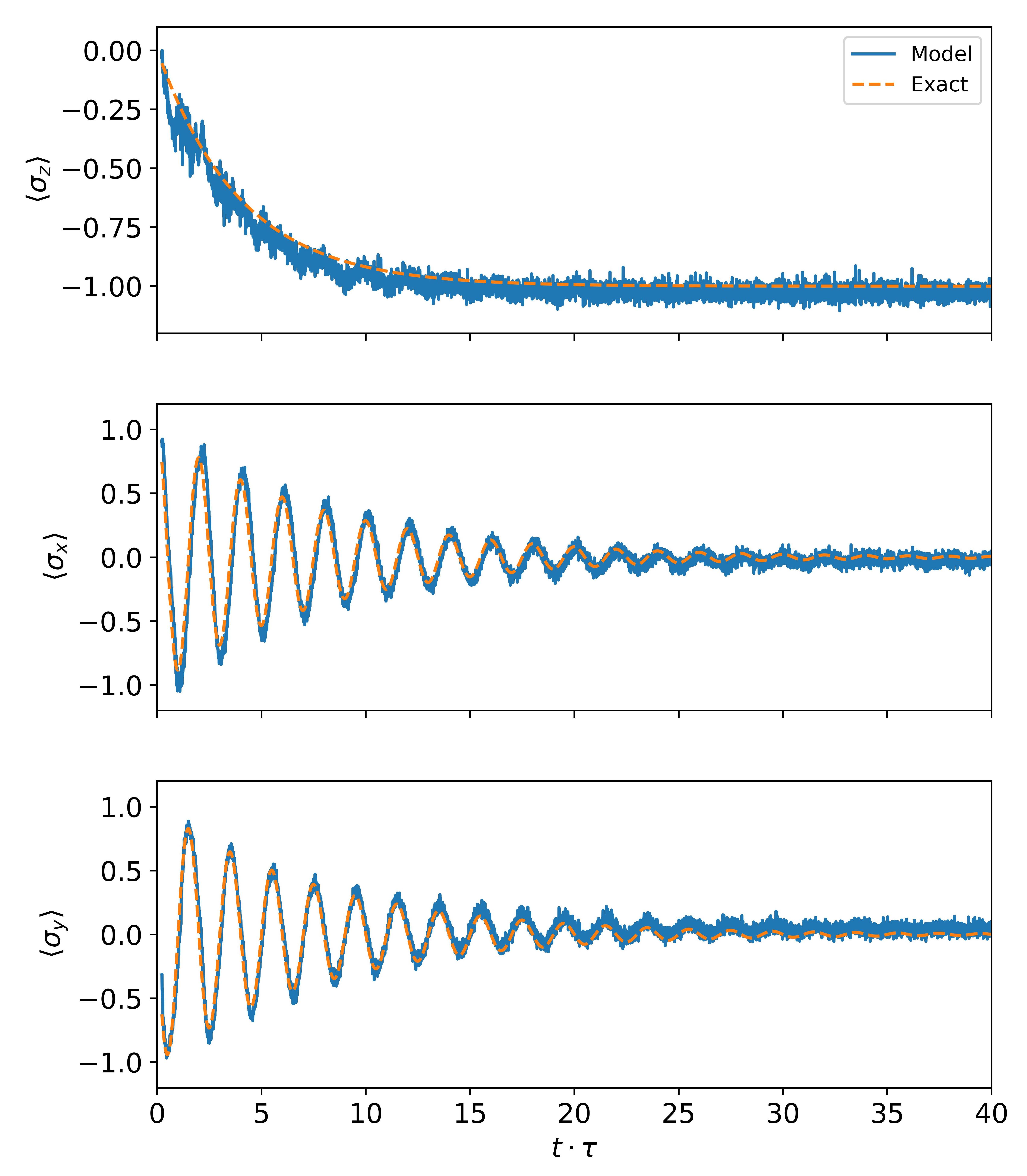}}
		\caption{40 units time expectation value $\langle\sigma_{z}\rangle$, $\langle\sigma_{x}\rangle$ and $\langle\sigma_{y}\rangle$. The coupling  intensity  $\gamma = 0.5$ and the initial state is $\ket{+}(\langle\sigma_{x}\rangle = 1)$. }
		\label{fig:2}
	\end{figure}

	To evaluate the feasibility of our method, we apply TSP to a dissipative model with Hamiltonian $H=g\sigma_{z}$ and Lindblad operator $L =\sigma_-=\frac{1}{2}\left(\sigma_{x}-i\sigma_{y}\right)$. We calculate the expectation values of the three Pauli operators $\sigma_z,\sigma_x$ and $\sigma_y$. In the long-time limit, the system will reach an equilibrium steady state. Specifically, we use the exact data which has been obtained by the Quantum Toolbox in Python \cite{johansson2012qutip} to make the prediction. The length of the time series will have directly impact on the the effects of the model's prediction. We find that the best result is achieved with selection of 240 samples per unit time. We construct a database by capturing the evolution of the reduced density matrix within the first 10 units of time for training the model. It is tested to find that our model can capture the trends and features of the data most efficiently only when the sequence length $L$ is set to 30. If the series length is too short, the time series data lacks distinguishable characteristics, and if the series length is too long, the model struggles to extract the features of the time series data.

\begin{figure}
		\centerline{\includegraphics[width=1\columnwidth]{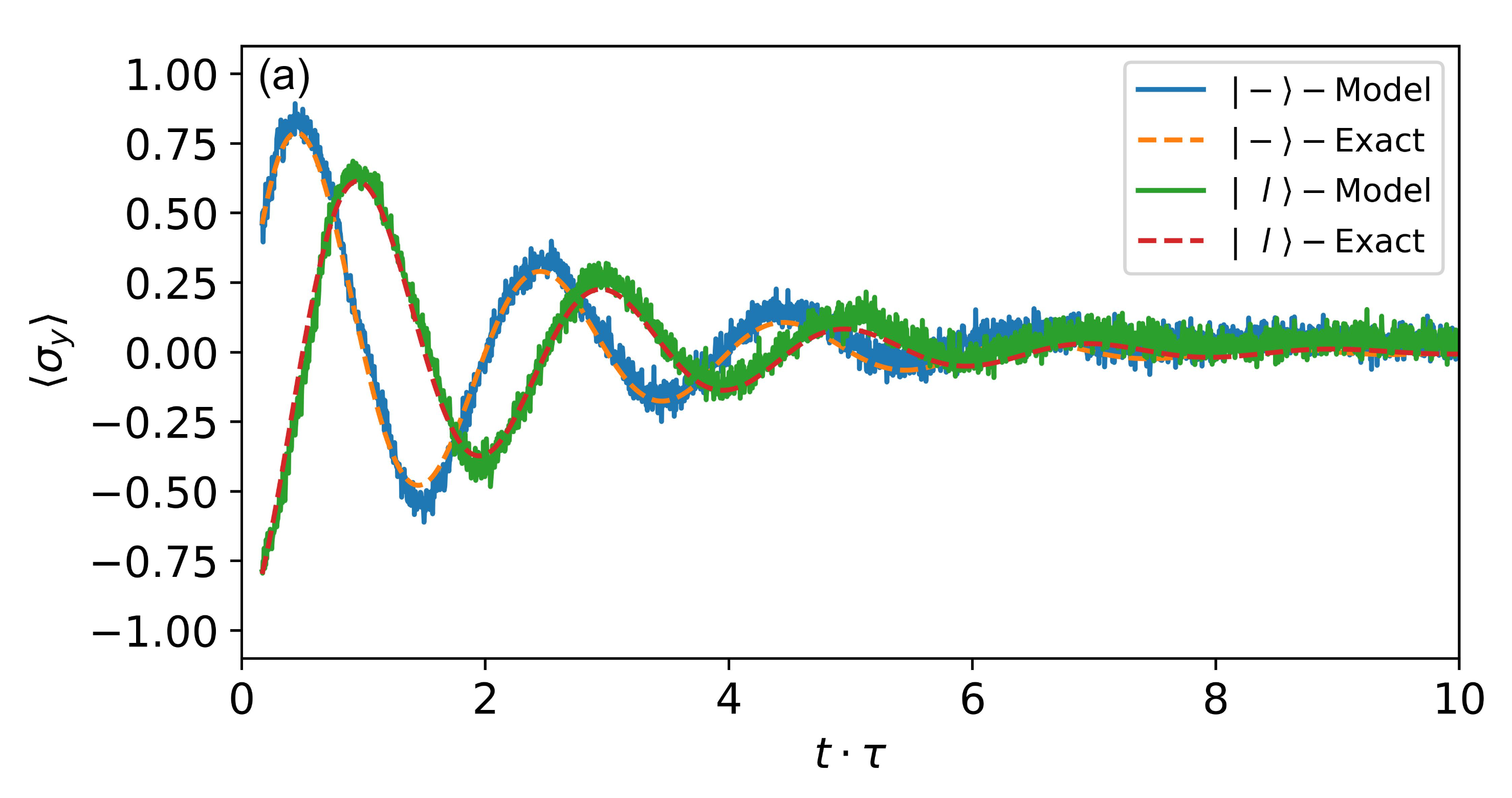}}
		\centerline{\includegraphics[width=1\columnwidth]{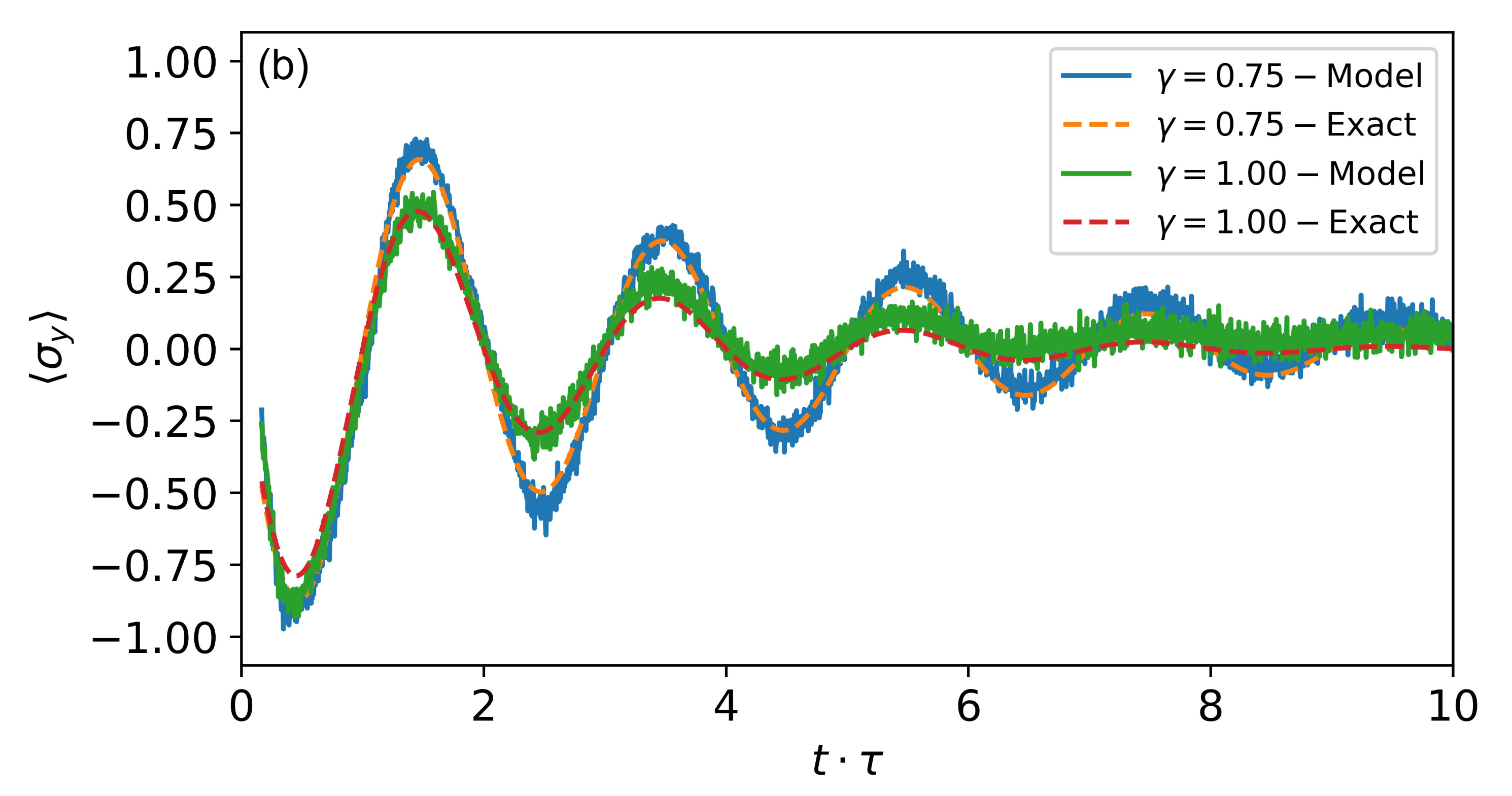}}
		\caption{The expectation value $\langle\sigma_{y}\rangle$ as a function of time. (a) The different initial states are  $\left| l \right \rangle (\langle\sigma_{y}\rangle = -1) $ and $\left|-\right\rangle (\langle\sigma_{x}\rangle = -1)$ .  The coupling  intensity $\gamma = 1$. (b) The different coupling intensities are $\gamma = 0.75$ and $\gamma = 1$. The initial state is $\ket{+}$.}
		\label{fig:3}
	\end{figure}

	In Fig.~\ref{fig:2}, we compare the time evolution of $\langle\sigma_z\rangle,\langle\sigma_x\rangle$ and $\langle\sigma_y\rangle$ between exact value and the prediction value using 40 units of time data. Fig.~\ref{fig:2} shows that the model can accurately predict the dynamical evolution of the open system within the first 10 unit time that the model has been trained. It can also give predictions in line with the changing trend when the average value decreases and tends to steady state in the long time limit. We also test our model with different initial states and coupling strengths, as shown in Figs.~\ref{fig:3}(a) (b). The results shows again that the trained model performs well under different conditions, although the margin of error becomes greater with time evolution. So our model has learned the characteristic of the evolution and has the ability to predict the future data based on past data. It is worth mentioning that we solely utilized the database for training, indicating the universality of our training strategy.

	\begin{figure}
		\centerline{\includegraphics[width=1\columnwidth]{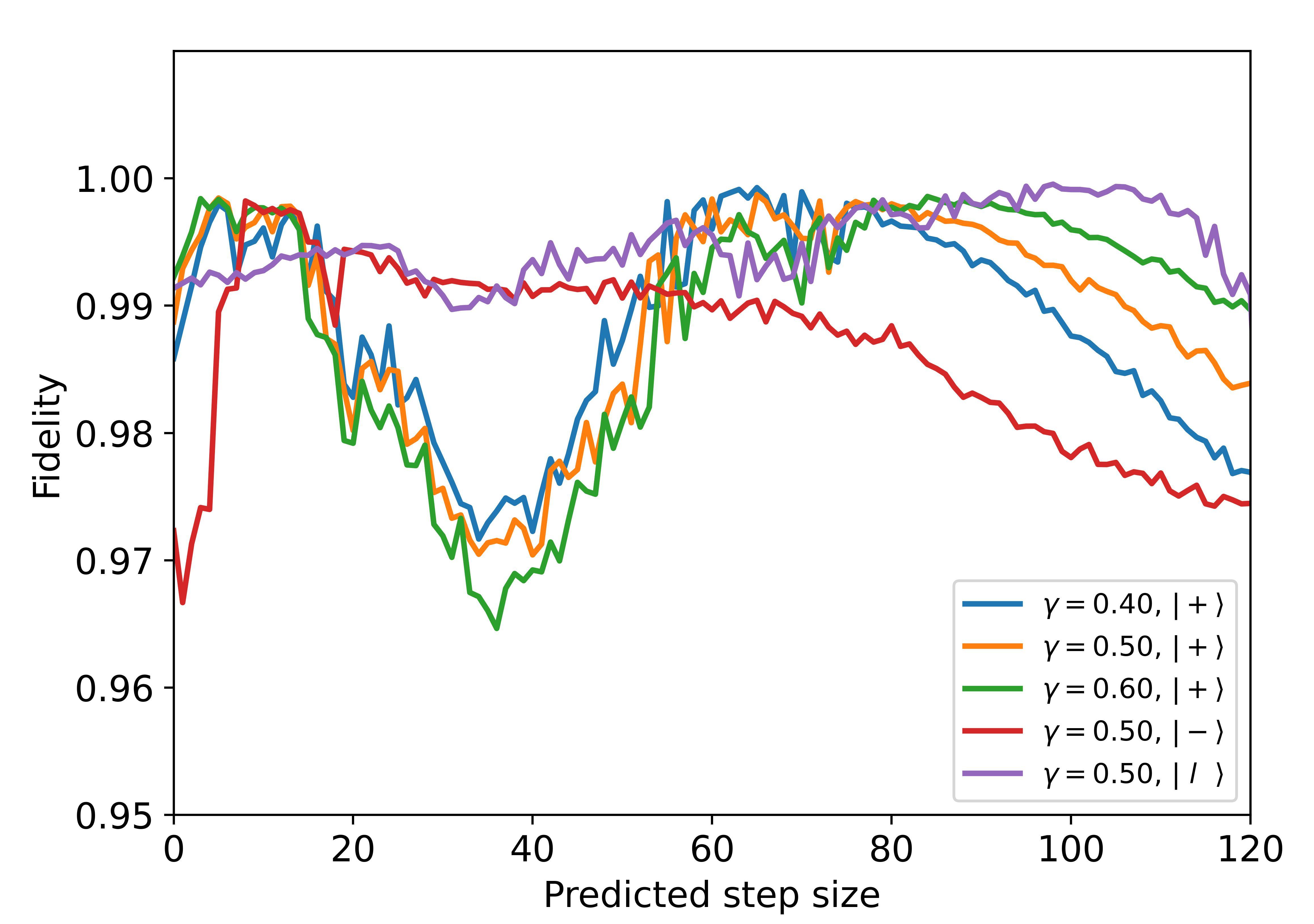}}
		\caption{Average fidelity of density matrix for the long-term TSP under different initial states and different coupling intensities.}
		\label{fig:4}
	\end{figure}

	\subsection{Long-term TSP}

	Next, we reorganiz the time series using the predicted data, as depicted in Fig.~\ref{fig:1}, to make further predictions for the future. When the prediction step exceeds the length of the time series, the predicted data completely replaces the original time series, and the model no longer relies on the original time series, that is, the long-term TSP. We select data from 30 time nodes starting from the initial moment to form our original time series. Using the long-term TSP strategy, we predict the reduced density matrix for the next 120 moments under different initial states and coupling intensities. The fidelity between the predicted density matrix and the exact density matrix is calculated by $F(\rho_1,\rho_2)=\text{Tr}(\sqrt{\sqrt{\rho_1}\:\rho_2\:\sqrt{\rho_1}})$. Here $\rho_{1,(2)}$ is the  predicted (exact) density matrix, respectively. In Fig.~\ref{fig:4} we plot the fidelity versus the predicted step size for different initial states and coupling intensities for long-term prediction. From Fig.~\ref{fig:4}, the fidelity exceeds 0.96 for all cases, indicating the accuracy prediction ability of our trained model. Note that the fidelities for different coupling intensities have their minimum value when the number of steps reaches approximately 40, the reason is that the reduced density matrix undergoes a significant change at this point. We can also see that as the number of prediction steps increases, the cumulative prediction error also increases as expected.

\begin{figure}
		\centerline{\includegraphics[width=1\columnwidth]{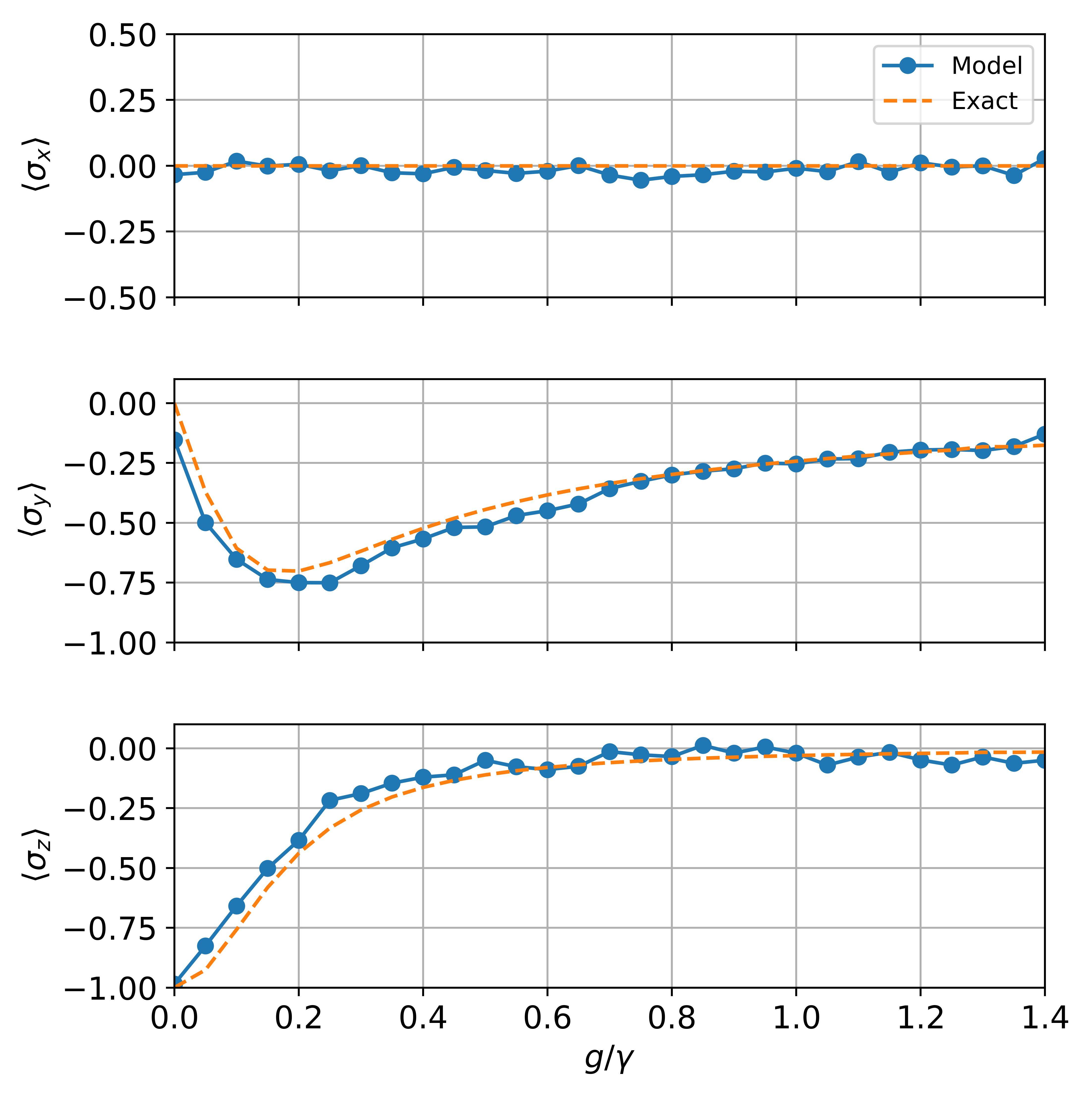}}
		\caption{The expectation value $\langle\sigma_{x}\rangle$,$\langle\sigma_{y}\rangle$ and $\langle\sigma_{z}\rangle$ for an open quantum system to reach a steady state at different $g/\gamma$. The Hamiltonian $H=g\sigma_{x}$ and the Lindblad operator $L =\sigma_- $.}
		\label{fig:5}
	\end{figure}

    To test the generality of our approach, now we turn to another model with  $H=g\sigma_{x}$ and Lindblad operator $L =\sigma_- $.  We will test the steady state value \cite{cui2015variational} that the model predicts to verify its performance. Fig.~\ref{fig:5} shows the expectation value of the Pauli operator when the system reaches steady state for different $g/\gamma$. In this case, we also use long-term TSP and choose 30 to 60 moments as our time series data. Then, we predict the data at each moment in turn until the steady state moment is obtained. A strong match between the model and the exact value can be found. Then our training method can be applied to different systems and the prediction is still effective in the long-term limit.

    \section{CONCLUSION}
    
  	In conclusion, we have developed a deep learning-based TSP model to predict the dynamics of open quantum systems. For both short-term and long-term TSP, we find that the trained model has the ability to predict the dynamics with high fidelity. The trained model also demonstrates scalability and performs well across different initial states and coupling intensities. Moreover, our train approach can be generalized to various quantum systems. It is necessary to make in-depth research on the prediction fidelity enhancement, especially the study of long-term prediction. Other TSP models \cite{tang2021building,grigorievskiy2014long,menezes2008long} might be used to achieve better results. Additionally, in this paper we only consider a simple single-qubit system. For multi-qubit systems, our method might lose its effectiveness due to the challenges associated with the high-dimensional probability distribution parameters after POVM measurement. In this case, the number of parameters scale exponentially ($ 4^N $) with the number of qubits, as a result the data characteristics of all parameters  are difficult to be captured simultaneously. Long-term prediction and extention to multi-quibt system will be the center of future study.

	\section{ACKNOWLEDGMENTS}
	This paper is supported by the Natural Science Foundation of Shandong Province (Grant No. ZR2021LLZ004) and Fundamental Research Funds for the Central Universities (Grant No. 202364008).

	\bibliographystyle{unsrt}
	\bibliography{reference.bib}
	
\end{document}